\documentclass[orivec,runningheads]{llncs}
\usepackage{amsfonts}
\usepackage{graphicx} 
\usepackage[T1]{fontenc}
\usepackage{amsmath}
\usepackage[utf8]{inputenc}
\usepackage{url}
\usepackage{listings}
\usepackage{algorithm}
\usepackage{algorithmicx}
\usepackage{algpseudocode}
\usepackage{multirow}
\usepackage{makecell} 
\usepackage{colortbl}
\usepackage{longtable}
\usepackage{multirow}
\usepackage[table]{xcolor}
\usepackage{float}
\usepackage[table]{xcolor}
\usepackage{booktabs}
\usepackage{tcolorbox}
\usepackage{tabularx}
\usepackage{array}
\usepackage[normalem]{ulem}

\definecolor{babypink}{RGB}{255, 228, 235}
\definecolor{pinkheader}{RGB}{255, 192, 203}

\usepackage{hyperref}
\hypersetup{
    colorlinks,
    linkcolor={red!50!black},
    citecolor={blue!50!black},
    urlcolor={blue!80!black}
}
\usepackage{cite}  
\usepackage{array} 
\usepackage{comment}
\title{Arc-Length Parameterized \mbox{Interpolating Splines}}
\author{
    Dafna K. Matsegora
    \and
    Stephen M. Watt
}
\institute{
  Cheriton School of Computer Science, University of Waterloo\\
  Waterloo, Ontario, Canada N2L 3G1\\
  \email{\{dafna.matsegora,stephen.m.watt\}@uwaterloo.ca}
}
\date{\today}
\begin{document}

\maketitle
\begin{abstract}
    We present an iterative algorithm to compute an arc-length parameterized spline interpolating a set of points.
    This differs from other methods where the computed spline either does not interpolate the original points or the parameterization is not the arc-length of the returned curves.
    Our method is applicable in any dimension $D \ge 2$, and we illustrate it with numerical results for plane curves.
\end{abstract}
\section{Introduction}
\label{sec:Introduction}

A common task in geometric modeling and its applications, such as path planning, computer-aided design and handwriting recognition, is the construction of smooth curves that interpolate a specific set of points. A standard approach is to use splines with some parameter, $t$ that increases monotonically along the curve. 
In many cases it is desirable to place some requirements on the parameterization.   

General parameterizations introduce a “speed function”, $||r'(t)||$, which leads to non-uniform traversal, inconsistent evaluation of geometric quantities, and added complexity in subsequent computations. As a result, many geometric and analytical quantities depend not only on the shape of the curve, but also on how it is parameterized. Arc-length reparameterization replaces the arbitrary time parameter with one that reflects only the curve's geometry, making the result independent of how fast or unevenly the curve is traced.

Parameterization by arc-length offers several advantages, as it significantly simplifies many mathematical expressions. In particular, various quantities become geometrically intrinsic, depending only on the shape of the curve. Furthermore, quantities defined per unit length become simpler to evaluate, whereas a general parameterization requires an additional Jacobian factor. This also leads to improved numerical stability, better-conditioned derivatives, and simpler curvature computation.

We are specifically interested in the problem of constructing a spline interpolant for a given set of  fixed points, such as might be obtained by sensor readings or geometric constraints. One easy parameterization is to join consecutive points with line segments and take the line segment arc-length as the parameter for an interpolating spline.  However, this parameterization lacks the desirable properties of using true arc-length.

This raises a natural practical question: given a fixed set of data points, how can one obtain an accurate arc-length parameterized interpolating spline? We propose an iterative algorithm that produces a close approximation in practice, achieving low error within just two iterations for most inputs, offering a fast and reliable approach to a problem for which no direct closed-form construction is generally available. Importantly, our problem is one of interpolation: the curve is required to pass exactly through the given sampled points, rather than approximating them or moving them as in many smoothing-based methods. 
This rules out smoothing or curve-approximation methods where control points can be adjusted to minimize error.
\section{Previous Work}
\label{sec:Previous Work}
Much existing work focuses on how arc-length-parameterized splines are applied in various settings, rather than on the underlying methods to compute them. The work reviewed in this section instead focuses specifically on existing approaches to arc-length-parameterized splines, grouped into four main categories: resampling and refitting, inverse-map composition, closed-form approximation, and iterative techniques. 

One approach~\cite{wang2002arc} takes a resampling-and-refitting approach: it computes the total arc length, places equally spaced points along the original spline, and constructs a new spline through those points, which are ultimately not the same as the sampled points unless the initial curve was linear. This method is well-suited to real-time simulation because the expensive steps can be carried out offline, making online evaluation fast. However, this is only advantageous when the curve is known in advance; in settings where the curve must be processed as it is generated, those preprocessing steps move online and can become too expensive. In addition, the method only approximates arc-length parameterization, since the spline is built from resampled points and it may not interpolate the given points.

Other work~\cite{peterson2006arc} describes an inverse-mapping composition approach: given an original spline $Q(u)$, it computes the cumulative arc-length function $L(u)$, solves $L(u)=a$ using Newton iteration to obtain sample inverse pairs $(a,u)$, fits an auxiliary spline $l(a) \approx L^{-1}(a)$, and forms the reparameterized curve $P(a)=Q(l(a))$. This preserves the original curve geometry exactly while 
enabling fast arc-length evaluation via spline composition. Peterson also notes that $Q(u)$ and $l(a)$ can be composed into a single higher-order curve $P(a)=Q(l(a))$, eliminating the need to evaluate two splines separately. However, the degree of $P$ grows as the product of the degrees of $Q$ and $l$, so the number of control points required grows very quickly, making this approach less practical for curves of even moderate complexity. Like resampling methods, it requires expensive preprocessing, which limits its use when curves must be reparameterized on-the-go. 


Another paper~\cite{giltextordmasculine1997new} also pursues an inverse-map approach, constructing piecewise low-degree approximations to $t(s)$ on each interval $[s_i, s_{i+1}]$ from sampled $(t_i, s_i)$ pairs. It avoids the need for higher-degree polynomials in regions of complex behaviour by working locally on each interval rather than fitting a single approximation over the full curve. However, like other inverse-map methods, the approach still requires a preprocessing step to compute arc-length samples before the reparameterization can be evaluated.

This inverse-mapping approach has been widely adopted in practice, as seen in implementations such as FullNitrous~\cite{fullnitrous2024}, which approximates the inverse via integrated Chebyshev polynomials, and the splines.readthedocs library~\cite{splines2023} using numerical quadrature and 
secant root-finding.

Another paper~\cite{walter1996approximate} takes a different approach altogether, replacing the lookup-table schemes used by other methods with an approximate closed-form solution. Instead of inverting the arc-length map numerically, they approximate the forward map $A(t)$, which relates the parametric variable to accumulated arc length, with a one or two-span B\'{e}zier curve. This avoids the need for a lookup table and allows arc length to be evaluated directly for any parameter value. The trade-off is reduced accuracy, making it better suited for applications that prioritize speed over precision.

Finally, another paper~\cite{hernandez2003sampling} proposes an iterative approach, \texttt{UniArcLength}, for re-parameterizing an explicit parametric curve $c(u)$ so that $n+1$ points are distributed uniformly with respect to arc-length. Their method constructs a monotonic $C^1$ spline $f^j(s)$ approximating the inverse arc-length function $\Phi(s)$, using both interpolation and derivative conditions at each iteration, and terminates when the computed arc-lengths fall within $\varepsilon$ of a fixed uniform target. While our method is also iterative, the two algorithms address fundamentally different problems. Their method requires an explicit parametric curve, which makes the inverse arc-length function $\Phi(s)$ accessible and allows derivative conditions to be enforced. In our setting, only a finite set of points is available: the underlying curve is unknown, $\Phi(s)$ cannot be evaluated, and there is no fixed uniform target to converge toward.

We have not found methods that place primary emphasis on the original data points.

\section{Iterative Method}
\label{sec:Iterative Method}

Our approach contrasts with the previous work by computing a sequence of splines iteratively.
At each stage, we construct an interpolating spline through the given points and iteratively refine the arc-length parameters until they are consistent with the spline's arc length. 

We begin with a parameterization based on a piecewise linear interpolation, assigning the chord length to each segment and cumulative chord length sum $s^{(0)}_i$ as the parameter value at each point $p_i = (x_i, y_i)$.
We then construct a spline of the desired degree through the given points using these parameter values.
New cumulative arc-lengths $s^{(1)}_i$ to the points are then calculated by numerical integration of the spline.
This is repeated  for a fixed number of iterations or until convergence of the $s^{(k)}_i$ values.

More precisely, given a set of points $\{p_i = (x_i, y_i)\}_{i=0}^{n}$, the initial 
estimate $s^{(0)}_i$ of the parameter for each point $p_i$ is the cumulative chord length up to that point. 
The total chord length is given by
\begin{equation*}
    L_{\text{chord}} = \sum_{i=1}^{n} \|p_i - p_{i-1}\|,
\end{equation*}
so that the initial parameter values are
\begin{equation*}
    s^{(0)}_i = \sum_{j=1}^{i} \|p_j - p_{j-1}\|, \quad i = 1, \ldots, n,
\end{equation*}
with $s^{(0)}_0 = 0$ and $s^{(0)}_n = L_{\text{chord}}$. 

At iteration $k$, coordinate splines $x^{(k)}(s)$ and $y^{(k)}(s)$
are constructed through the points using the $s^{(k)}_i$ values as
the parameter values. The arc length of the $i$th piece of the
resulting spline is then computed by numerically integrating
\[
    \ell^{(k)}_i =
    \int_{s^{(k)}_i}^{s^{(k)}_{i+1}}
    \sqrt{
       \left(\frac{dx^{(k)}}{ds}\right)^2 +
       \left(\frac{dy^{(k)}}{ds}\right)^2
    }\,ds .
\]
The next parameter values are obtained as the cumulative arc lengths,
\[
    s^{(k+1)}_i = \sum_{j=0}^{i-1} \ell^{(k)}_j,
    \quad i = 0,1,\ldots,n .
\]
This process is repeated either for a fixed number of iterations, when speed is the primary concern, or until the $s^{(k)}_i$ values converge.

Above we have described the process for curves in the plane, that is for dimension $2$.  For dimension $D$, we have $p_i = (x^1_i, \ldots, x^D_i)$, $D$ coordinate splines are computed and
\begin{equation*}
    \ell^{(k)}_i = \int_{s^{(k)}_i}^{s^{(k)}_{i+1}} \sqrt{~\sum_{j=1}^D \left(\frac{dx^{j,(k)}}{ds}\right)^2~ } \, ds.
\end{equation*}
Thus the parameter values at the interpolation points are made consistent
with the arc lengths of the corresponding spline pieces; no separate
inverse-map reparameterization is introduced within each piece.
Note our method is agnostic to the degree of the interpolating splines.


\subsection{Input Data and Point Distribution}
\label{Input Data and Point Distribution}

The algorithm operates on any fixed set of ordered data points, regardless of how they were collected. When points are synthetically generated or placed at uniform intervals, the method applies without modification. Often, however, data points are fixed sensor measurements collected from a digitizing device such as a stylus or e-pen, sampled at a fixed temporal frequency. In either case, the points are not adjustable and cannot be moved or resampled without altering the data.

Additionally, since sampling is time-based rather than geometry-based, the distribution of points along the curve is often inherently non-uniform with respect to arc length. In particular, humans naturally slow down near sharp turns and corners, resulting in a higher density of sampled points in those regions. 
This dependence of the sample distribution on writing speed is one of the motivations for our work: without arc-length reparameterization, two people writing the same shape at different speeds may produce different parameter functions for geometrically similar traces.

\begin{figure}[htbp]
    \centering
    \includegraphics[width=5cm]{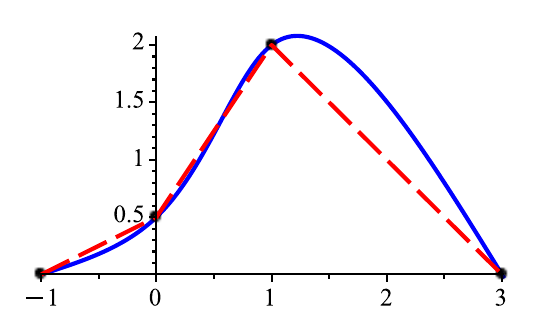}
    \caption{Spline behaviour near a geometric corner. The interpolating spline produces a smooth approximation rather than preserving the sharp corner exactly.}
    \label{fig:spline-corner}
\end{figure}

Near corners and near-singular points, the spline will produce a smooth approximation, since splines are differentiable everywhere by construction. As shown in Figure~\ref{fig:spline-corner}, true geometric corners are not representable exactly, and the method implicitly approximates them with a smooth curve through the locally dense cluster of points. Consequently, regions with corners or rapid changes in curvature often require more iterations to converge. 
The arc length computed by the method is therefore the arc length of the smooth interpolating spline, not of an idealized non-smooth curve with an exact sharp corner.

\subsection{Spline Representation}

For our experiments, we used cubic splines, which offer a practical balance
between computational cost and smoothness. Cubic splines provide $C^2$
continuity, matching both first and second derivatives across knots. This is
useful for producing smooth interpolants and for later geometric quantities
such as curvature, while keeping the numerical arc-length computation
relatively inexpensive. Higher-degree splines would increase computational cost with very little meaningful improvement in the accuracy of the arc-length estimates in practice.

\begin{lstlisting}[float,caption={Maple implementation of the arc-length reparameterization algorithm}, label={lst:full},
%xleftmargin=-8mm,
%xrightmargin=-8mm,
columns=fullflexible,
basicstyle=\fontsize{9}{10}\selectfont\ttfamily
%basicstyle=\ttfamily\footnotesize
]
Hypot := (p1,p0) -> sqrt((p1[1] - p0[1])^2 + (p1[2] - p0[2])^2);
LastOp      := v -> op(nops(v),v);
IsPiecewise := f -> evalb(type(f,'function') and op(0,f) = 'piecewise');

PiecewiseInterpFns := proc(pw)
    local i; [seq(op(2*i,pw), i = 1..iquo(nops(pw),2)), LastOp(pw)];
end proc;

CumulativePiecewiseLinearLengths := proc(xydata)
    local i, lengths, len; len := 0; lengths := [len];
    for i from 2 to nops(xydata) do
        len := len + evalf(Hypot(xydata[i],xydata[i - 1]));
        lengths := [op(lengths),len];
    end do;
    lengths
end proc;

ArcLengthOfPlaneCurve := proc(xfun,yfun,t,tStart,tEnd)
    local dxdt,dydt,integrand; dxdt:= diff(xfun,t); dydt:= diff(yfun,t);
    integrand := sqrt(dxdt^2 + dydt^2);
    evalf(Int(integrand,t = tStart..tEnd))
end proc;

CumulativePiecewiseFnLengths := proc(xfun,yfun,v,vvals)
    local xparts, yparts, i, len, lengths;
    xparts := PiecewiseInterpFns(xfun); 
    yparts := PiecewiseInterpFns(yfun);
    len := 0; lengths := [len];
    for i to nops(xparts) do
        len := len + ArcLengthOfPlaneCurve(xparts[i],yparts[i],v,
                                               vvals[i],vvals[i+1]);
        lengths := [op(lengths),len];
    end do;
    lengths
end proc;

IterateLengths := proc(xyvals,endptsVal,epsilon)
    local s, svals, svals_prev, xvals, yvals, xspline, yspline, 
    iter, max_change, i;
    svals := CumulativePiecewiseLinearLengths(xyvals);
    xvals := map(p -> op(1,p),xyvals); yvals := map(p -> op(2,p),xyvals);
    iter := 0; max_change := infinity;
    while epsilon < max_change do
        iter := iter + 1; svals_prev := svals;
        xspline := Spline(svals,xvals,s,degree = 3,endpoints = endptsVal);
        yspline := Spline(svals,yvals,s,degree = 3,endpoints = endptsVal);
        svals := CumulativePiecewiseFnLengths(xspline,yspline,s,svals);
        max_change := max(seq(abs(svals[i]-svals_prev[i]),
                                i = 1..nops(svals))) / svals[-1];
    end do; 
    [xspline, yspline, LastOp(svals)]
end proc;
\end{lstlisting}

\subsection{Convergence Criterion}
Two observations about the algorithm's behaviour are worth noting. First, the initial chord-length estimate is generically an underestimate: a straight line between two points is never longer than the curve connecting them, so the algorithm begins with a systematic bias that the iterations 
correct. Second, the convergence rate depends on the curve's geometry. Smooth, symmetric curves, such as a circle, tend to converge in very few iterations, since arc-length corrections are distributed uniformly across all segments. Curves with regions of high curvature or near-corners require more iterations, as the arc-length error is concentrated locally and redistributes more slowly. In practice, we find that two iterations are sufficient for most inputs to achieve low error.

When running until convergence, the algorithm terminates when the maximum change in the $s_i$ values, normalized by the total arc-length, falls below a prescribed tolerance $\varepsilon$:
\begin{equation*}
    \frac{\max_{i} \, |s_i^{(k+1)} - s_i^{(k)}|}{s_n^{(k+1)}} < \varepsilon.
\end{equation*}
Normalizing by $s_n^{(k+1)}$ makes the criterion scale-invariant, so that $\varepsilon$ has a consistent meaning regardless of the overall size of the curve. In 
practice, we find that two iterations are sufficient for most inputs to achieve low normalized error.

\renewcommand{\arraystretch}{1.05}

\begin{small}
\begin{table}
\centering

\caption{Iterations to convergence for different shapes across point counts $n$ and tolerances $\varepsilon$, the maximum relative segment arc-length change between iterations.}
\label{tab:tablefull2}

\setlength{\tabcolsep}{6.2pt}
\begin{tabular}{
p{2.5cm}
>{\centering\arraybackslash\columncolor{pink!10}}p{0.9cm}
*{10}{>{\columncolor{pink!10}}c}
}

\hline
\rowcolor{pink!40}
\centering{Shape} & $\log_{10}(\varepsilon)$ 
& 12 & 24 & 36 & 48 & 60 & 72 & 84 & 96 & 108 & 120 \\ \hline

\centering\multirow{6}{*}{\shortstack{{Ellipse}\\
\includegraphics[width=2cm]{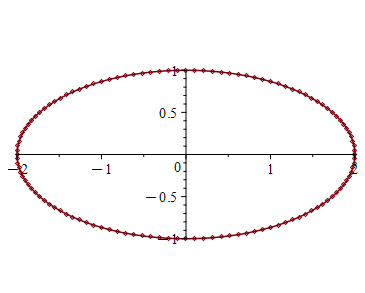}}}

& -1 & 1 & 1 & 1 & 1 & 1 & 1 & 1 & 1 & 1 & 1 \\ 
& -2 & 2 & 2 & 1 & 1 & 1 & 1 & 1 & 1 & 1 & 1 \\ 
& -3 & 2 & 2 & 2 & 1 & 1 & 1 & 1 & 1 & 1 & 1 \\ 
& -4 & 2 & 2 & 2 & 2 & 2 & 2 & 2 & 2 & 2 & 2 \\ 
& -5 & 2 & 2 & 2 & 2 & 2 & 2 & 2 & 2 & 2 & 2 \\ 
& -6 & 2 & 2 & 2 & 2 & 2 & 2 & 2 & 2 & 2 & 2 \\  \hline

\centering\multirow{6}{*}{\shortstack{{Rose}\\
\includegraphics[width=2cm]{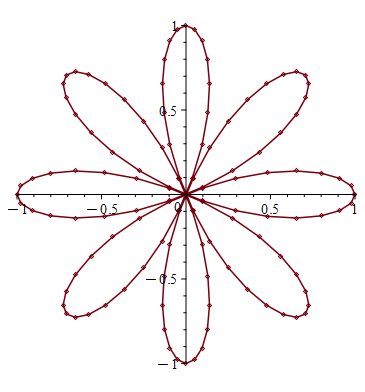}}}

& -1 & 1 & 1 & 1 & 1 & 1 & 1 & 1 & 1 & 1 & 1 \\
& -2 & 3 & 2 & 2 & 2 & 2 & 1 & 1 & 1 & 1 & 1 \\
& -3 & 4 & 3 & 3 & 2 & 2 & 2 & 2 & 2 & 2 & 2 \\
& -4 & 6 & 4 & 5 & 3 & 2 & 2 & 2 & 2 & 2 & 2 \\
& -5 & 8 & 5 & 6 & 3 & 4 & 3 & 2 & 3 & 2 & 2 \\
& -6 & 9 & 6 & 8 & 4 & 5 & 4 & 3 & 3 & 3 & 3 \\ \hline

\centering\multirow{6}{*}{\shortstack{{Heart}\\
\includegraphics[width=2cm]{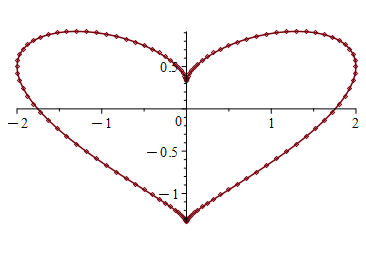}}}

& -1 & 1 & 1 & 1 & 1 & 1 & 1 & 1 & 1 & 1 & 1 \\
& -2 & 2 & 2 & 1 & 1 & 1 & 1 & 1 & 1 & 1 & 1 \\
& -3 & 3 & 2 & 2 & 2 & 2 & 2 & 1 & 1 & 1 & 1 \\
& -4 & 4 & 2 & 2 & 2 & 2 & 2 & 2 & 2 & 2 & 2 \\
& -5 & 6 & 3 & 3 & 2 & 2 & 2 & 2 & 2 & 2 & 2 \\
& -6 & 7 & 4 & 3 & 3 & 3 & 3 & 2 & 2 & 2 & 2 \\  \hline

\centering\multirow{6}{*}{\shortstack{{Infinity}\\
\includegraphics[width=2cm]{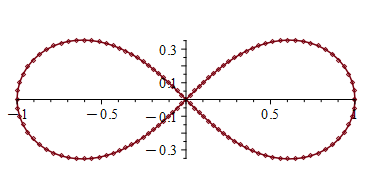}}}

& -1 & 1 & 1 & 1 & 1 & 1 & 1 & 1 & 1 & 1 & 1 \\
& -2 & 2 & 1 & 1 & 1 & 1 & 1 & 1 & 1 & 1 & 1 \\
& -3 & 2 & 2 & 2 & 2 & 2 & 2 & 1 & 1 & 1 & 1 \\
& -4 & 3 & 2 & 2 & 2 & 2 & 2 & 2 & 2 & 2 & 2 \\
& -5 & 4 & 2 & 2 & 2 & 2 & 2 & 2 & 2 & 2 & 2 \\
& -6 & 5 & 3 & 2 & 2 & 2 & 2 & 2 & 2 & 2 & 2 \\ \hline

\centering\multirow{6}{*}{\shortstack{{Hypotrochoid}\\
\includegraphics[width=1.5cm]{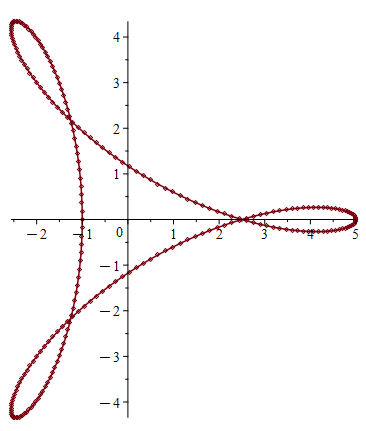}}}

& -1 & 1 & 1 & 1 & 1 & 1 & 1 & 1 & 1 & 1 & 1 \\
& -2 & 1 & 1 & 1 & 1 & 1 & 1 & 1 & 1 & 1 & 1 \\
& -3 & 2 & 2 & 2 & 1 & 1 & 1 & 1 & 1 & 1 & 1 \\
& -4 & 2 & 2 & 2 & 2 & 2 & 2 & 2 & 2 & 2 & 1 \\
& -5 & 3 & 3 & 2 & 2 & 2 & 2 & 2 & 2 & 2 & 2 \\
& -6 & 4 & 3 & 3 & 2 & 2 & 2 & 2 & 2 & 2 & 2 \\  \hline

\centering\multirow{6}{*}{\shortstack{{Lissajous}\\
\includegraphics[width=1.8cm]{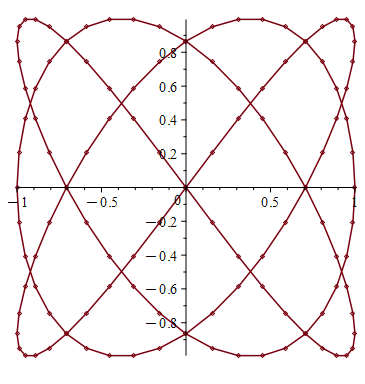}}}

& -1 & 1 & 1 & 1 & 1 & 1 & 1 & 1 & 1 & 1 & 1 \\
& -2 & 2 & 2 & 2 & 1 & 1 & 1 & 1 & 1 & 1 & 1 \\
& -3 & 2 & 3 & 2 & 2 & 2 & 2 & 2 & 2 & 2 & 2 \\
& -4 & 4 & 5 & 4 & 3 & 2 & 2 & 2 & 2 & 2 & 2 \\
& -5 & 6 & 6 & 5 & 4 & 3 & 2 & 2 & 2 & 2 & 2 \\
& -6 & 9 & 8 & 7 & 6 & 3 & 4 & 3 & 2 & 3 & 3 \\ \hline

\centering\multirow{6}{*}{\shortstack{{Astroid}\\
\includegraphics[width=2cm]{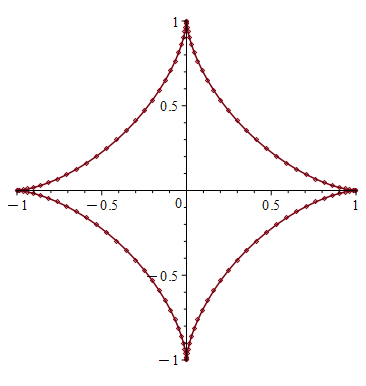}}}

& -1 & 1 & 1 & 1 & 1 & 1 & 1 & 1 & 1 & 1 & 1 \\
& -2 & 2 & 1 & 1 & 1 & 1 & 1 & 1 & 1 & 1 & 1 \\
& -3 & 3 & 2 & 2 & 1 & 1 & 1 & 1 & 1 & 1 & 1 \\
& -4 & 4 & 2 & 2 & 2 & 2 & 2 & 2 & 2 & 2 & 2 \\
& -5 & 5 & 3 & 2 & 2 & 2 & 2 & 2 & 2 & 2 & 2 \\
& -6 & 6 & 3 & 3 & 3 & 2 & 2 & 2 & 2 & 2 & 2 \\ \hline
\end{tabular}
\end{table}
\end{small}
\section{Experimental Results}
\label{sec:Experimental Results}
For the numerical experiments reported here, and in the Maple implementation
shown in Listing~\ref{lst:full}, we used cubic interpolating splines.
The iterative method itself is independent of this particular degree.

To evaluate the algorithm, we tested it on a diverse set of parametric curves, including: Circle, Ellipse, Rose, Heart, Infinity, Hypotrochoid (3), Astroid, and Lissajous. These shapes were chosen to cover a range of geometric behaviours, from smooth and symmetric curves such as the circle and ellipse, to curves with high curvature regions, cusps, and self-intersections.

For each shape, points were sampled at $n \in \{12k: k = 1, \dots, 10\}$ uniformly distributed parameter values, and the 
algorithm was run with convergence tolerances $\varepsilon \in \{10^{-1}, 10^{-2}, 10^{-3}, 10^{-4}, 10^{-5}, 10^{-6}\}$. All experiments were implemented and run in Maple (see Listing~\ref{lst:full}), using periodic end point conditions to construct a cubic spline at each iteration. Representative results are presented in Table~\ref{tab:tablefull2}, omitting the circle case.

\begin{figure}[htbp]
    \centering
    \includegraphics[width=11cm]{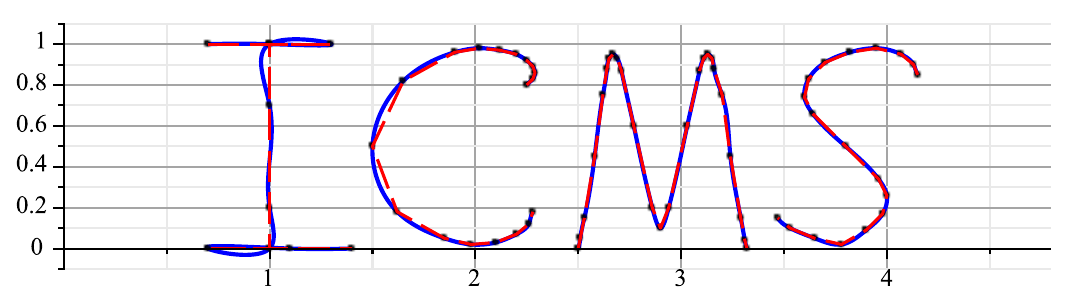}
    \caption{Spline representation (in blue) of the handwritten term 'ICMS' demonstrating the ability of the proposed method to approximate more complex handwriting data with multiple curvature changes and corner-like features.}
    \label{fig:spline-ICMS}
\end{figure}

To further demonstrate the applicability of the proposed approach to more realistic examples, a sample handwriting data set representing the term 'ICMS' was considered. The data were recorded as a sequence of $(x,y)$ coordinates. As discussed in Section~\ref{Input Data and Point Distribution}, handwritten symbols containing corners or near-singular points often exhibit a denser distribution of sampled data points. Figure~\ref{fig:spline-ICMS} shows the resulting spline representation of the handwriting sample.

\section{Conclusions and Future Work}
\label{sec:Future Work}

We have shown a direct method to compute arc-length parameterized splines through a given set of points.
As expected, the number of iterations depends on the number of sample points and the complexity of the resulting shape.  For the sorts of curves occurring in many applications, two iterations suffice.

The proposed method opens several directions for future work. One natural application is in handwriting recognition. Earlier work relies on chord-length parameterizations of piecewise linear curves. While these are simple to compute, they often do not capture the underlying smooth structure of handwritten input particularly well. Replacing these representations with arc-length-parameterized interpolating splines, which provide closer models of the written traces, is believed to improve classification accuracy, especially in handwritten mathematical expressions, where small geometric differences are important. 

Future work should compare this approach with existing preprocessing methods that use piecewise linear parameterizations. The main focus should be on how it affects classification accuracy and how robust it is to noise and variations in handwriting. It is also important to consider the computational cost of the iterative procedure to assess whether it is practical in real-world settings.

From an implementation perspective, further development in TypeScript, C++, and Maple would allow the method to be tested across different environments. A TypeScript implementation, in particular, would enable integration of the algorithm into browser-based tools for real-time handwriting input, where efficiency and responsiveness are critical.

Finally, while we have seen that the method converges rapidly in the examples studied, it remains to prove convergence and uniqueness for splines of the desired type.

\bibliographystyle{splncs04}
\bibliography{main}

\end{document}